\documentclass{INTERSPEECH2023}
\interspeechcameraready 

\usepackage{xcolor}
\usepackage{enumitem}
\usepackage{bm}
\usepackage{amssymb}
\usepackage{amsmath}
\usepackage{hyperref}
\usepackage{graphicx}
\usepackage{algpseudocode}
\usepackage{enumerate}
\usepackage{booktabs}
\usepackage{makecell}
\usepackage{threeparttable}
\usepackage{tabularx} 
\usepackage{multirow}
\usepackage{ragged2e}
\usepackage{url}
\usepackage{color}

\providecommand{\zd}[1]{\textcolor{black}{{#1}}}

\title{Phase perturbation improves channel robustness for speech spoofing countermeasures}
\name{Yongyi Zang, You Zhang, Zhiyao Duan}
\address{University of Rochester, Rochester, New York, USA}
\email{\{yongyi.zang, you.zhang, zhiyao.duan\}@rochester.edu}

\begin{document}

\maketitle
 
\begin{abstract}
In this paper, we aim to address the problem of channel robustness in speech countermeasure (CM) systems, which are used to distinguish synthetic speech from human natural speech. On the basis of two hypotheses, we suggest an approach for perturbing phase information during the training of time-domain CM systems. Communication networks often employ lossy compression codec that encodes only magnitude information, therefore heavily altering phase information. Also, state-of-the-art CM systems rely on phase information to identify spoofed speech. Thus, we believe the information loss in the phase domain induced by lossy compression codec degrades the performance of the unseen channel. We first establish the dependence of time-domain CM systems on phase information by perturbing phase in evaluation, showing strong degradation. Then, we demonstrated that perturbing phase during training leads to a significant performance improvement, whereas perturbing magnitude leads to further degradation.
\end{abstract}
\noindent\textbf{Index Terms}: speech recognition, human-computer interaction, computational paralinguistics

\section{Introduction}
Speech generative systems have been progressing rapidly in recent years~\cite{9262021, tan2021survey}. The state-of-the-art speech generative model VALL-E~\cite{wang2023neural} can even mimic a person's voice with only 3 seconds of speech. If misused by criminals, these deep generative algorithms could aid in spoofing attacks. Therefore, the research community has been developing speech countermeasure (CM) systems for distinguishing synthetic speech from human natural speech. For real-world applications, spoofing attacks could be generated by algorithms and transmitted through communication channels that are novel to the CM system. Therefore, CM systems need to generalize to unseen synthetic attacks and channel variations~\cite{zhang2023generalize}.

Such generalization ability has been studied by the CM community, especially driven by the ASVspoof challenge series~\cite{yamagishi21_asvspoof}. In ASVspoof2019, the spoofing attacks in the evaluation set are created with different generative algorithms than the training set, enforcing the evaluation on the generalization ability to unseen attacks. Raw-waveform-based CM systems demonstrated especially great performance compared to frequency-magnitude-based CM systems.
In ASVspoof2021~\cite{liuASVspoof2021Spoofed2022}, the issue of channel robustness was investigated by introducing channel variation on the evaluation set of ASVspoof2019. The results show strong performance degradation when evaluating on unseen channels. There have been some solutions provided by the participants of ASVspoof2021~\cite{benhafid21_asvspoof, chen21_asvspoof, chen21b_asvspoof, tomilov21_asvspoof} that mainly use empirical methods to mitigate the problem, such as data augmentation and feature engineering.

In this paper, we propose to alleviate the channel robustness issue of state-of-the-art CM systems through phase perturbation in training. This proposal is developed based on two observations and our hypotheses for explaining the observations: 1) State-of-the-art CM systems are time-domain systems; We hypothesize that they rely on phase information to detect synthetic spoofing attacks, and 2) Communication channels often employ lossy compression codecs that are designed to only encode magnitude information; We hypothesize that they alter phase information in speech, making phase-aware CM systems difficult to generalize to unseen channels. Based on these hypotheses, we propose to perturb the phase when training phase-aware CM systems. We believe that this will make such systems less reliant on phase information, hence more robust to channel variations.

We design a set of experiments to test our hypotheses. By training three state-of-the-art time-domain CM systems and \zd{perturbing} phase during evaluation, we discover that all CM systems' performance degrades as the phase perturbation amount increases. We also perturb magnitude during evaluation for comparison, and observe that better performing CM systems show a stronger degradation effect when the phase is perturbed compared to magnitude, suggesting that the performance boost in time-domain CM system is likely due to better utilization of phase information, \zd{yet such utilization becomes overfitting when phase is perturbed during evaluation}. 

Then, we employ the best performing CM system AASIST \cite{Jung2021AASIST} and perturb phase during training, and evaluate on data from unseen transmission channels. We observe a significant performance improvement, where the best configuration demonstrates a 26.2\% \zd{relative improvement on the equal error rate (EER). As a comparison,} magnitude perturbation during training degrades performance in all settings. \zd{This suggests that the lossy compression in channel effects indeed retains magnitude information but corrupts phase information, and the phase perturbation in training is an effective strategy toward mitigating the channel robustness issues of time-domain CM systems. To our best knowledge, this is the first study that investigates the effects of phase and magnitude perturbation on CM systems.} 
All code, audio samples, and trained model
are open-sourced\footnote{\url{https://yongyi.dev/phase-antispoofing}}.

\begin{figure*}[t]
  \centering
  \includegraphics[width=\textwidth]{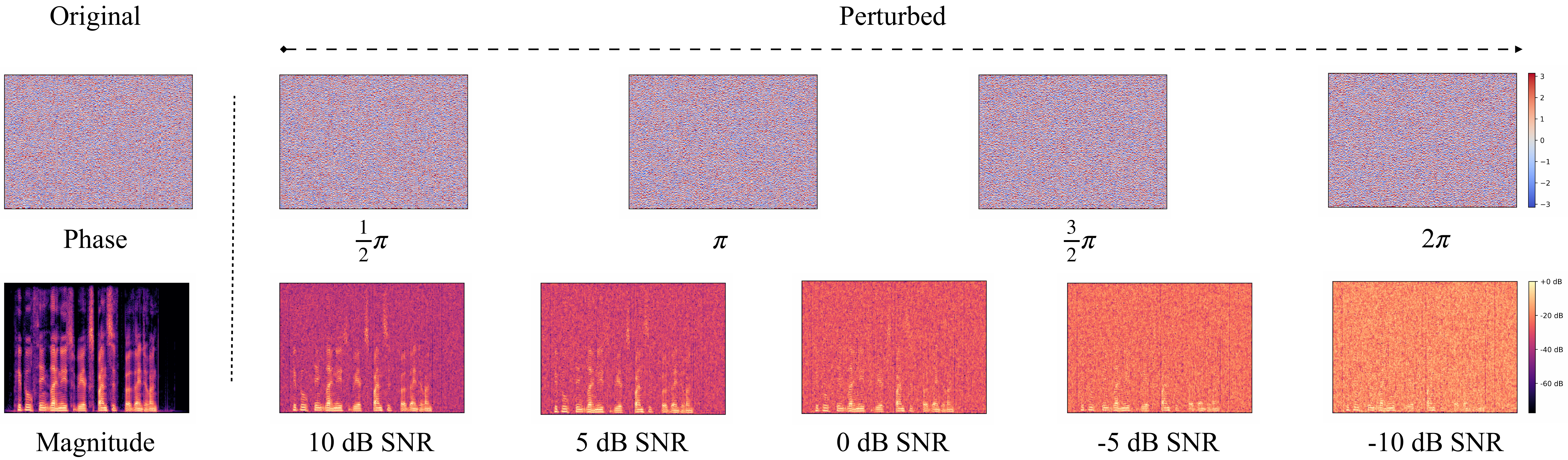}
\vspace{-10pt}
\caption{Magnitude and phase spectrograms with different amounts of perturbation, using utterance LA\_T\_1000137 as an example. 
The horizontal axis is time from 0 to 2.5 s, while the vertical axis is frequency from 0 to 8 kHz.}
\label{fig:perexp}
\end{figure*}

\section{CM systems rely on phase information}
In this section, we study the dependence of raw-waveform-based CM systems on phase information. We select three state-of-the-art CM systems with top performance on the ASVSpoof2019LA dataset \cite{yamagishi21_asvspoof}: RawNet-2 
 \cite{jung2019RawNet}, RawGAT-ST \cite{takEndtoEndSpectroTemporalGraph2021}, and AASIST \cite{Jung2021AASIST}. 


\subsection{Dataset}
We employ ASVspoof2019LA~\cite{wangASVspoof2019Largescale2020}, which is the logical access (LA) subset of the ASVspoof2019 challenge. It contains bonafide speech from the VCTK corpus \cite{veaux2017cstr}, with a vast variety of attacks. We follow the same training, development, and evaluation split as the ASVspoof2019 protocol. For the spoofing attacks, the training and development sets contain the same 6 attacks, and the evaluation set contains 13 unseen attacks.

\subsection{Experimental setup} 
\textbf{Phase perturbation.} During evaluation, we randomly perturb the phase at different amounts.
For an utterance with perturbation amount $n$, we extract its phase and magnitude spectrograms. The phase $\phi$ of every bin in the phase spectrogram is randomly reassigned a value in the range of $[\phi-\frac{n}{2}$, $\phi+\frac{n}{2}]$, while the magnitude spectrogram is unchanged. The phase perturbed utterance is then synthesized \zd{using} the original magnitude and the perturbed phase spectrograms. We select four perturbation settings, with phase perturbation $n$ as $\frac{1}{2}\pi$, $\pi$,$\frac{3}{2}\pi$ and $2\pi$, respectively.

\begin{figure}[]
  \centering
  \centerline{\includegraphics[width=\columnwidth]{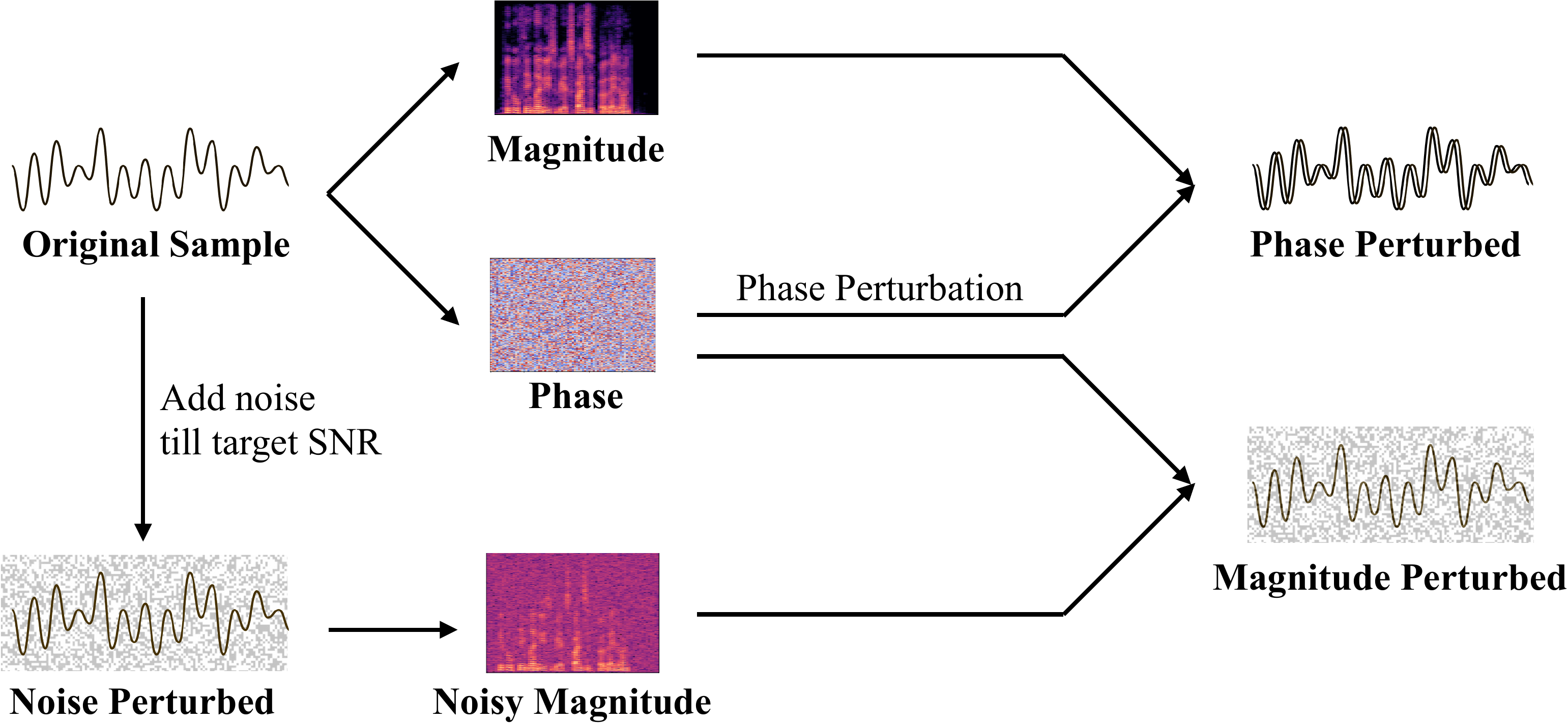}}
\vspace{-1pt}
\caption{Perturbation process for phase and magnitude corrupted data. For noise perturbation, the noise is directly added to the waveform.}
\label{fig:perturb}
\end{figure}

\textbf{Magnitude perturbation.} As a comparison, we also conduct a set of experiments with magnitude perturbed. For an utterance with a perturbation amount $m$ dB signal-to-noise ratio (SNR), we add white noise to the utterance \zd{to make its} SNR equal to $m$ dB, then extract its magnitude spectrogram. 
The magnitude-perturbed utterance is then synthesized from the noisy magnitude spectrogram and the original utterance's phase spectrogram. If the synthesized audio \zd{waveform exceeds} the maximum limit, we apply normalization to prevent clipping. We selected five perturbation settings for magnitude: 10 dB, 5 dB, 0 dB, -5 dB and -10 dB SNR. Figure~\ref{fig:perturb} visualizes both phase and magnitude perturbation processes, while Figure~\ref{fig:perexp} shows an example utterance with all perturbation settings. A setting without perturbation is also introduced as the baseline.

\textbf{Evaluation metric.} We follow the ASVspoof challenge series and employ equal error rate (EER), defined as the point where the false acceptance rate equals the false rejection rate. Lower EER indicates better performance. 

\textbf{Training details.} All time-domain CMs are trained with the ASVspoof2019LA training set, and validated on the development set. To reproduce the best performance for all CM systems, we use the hyper-parameter settings as reported in \cite{Jung2021AASIST}, and select the checkpoint with the lowest validation EER within 100 epochs for evaluation. All models are trained with three distinct random seeds and reported with an averaged result to mitigate the impact of the random seed on model performance~\cite{wangComparativeStudyRecent2021}.

\subsection{Results and analyses}
Table~\ref{tab:resdegphase} demonstrates the EER performance for all three CM systems evaluated with phase perturbation. As a sanity check, the performance of all CM systems on ASVspoof2019LA is similar to that reported in \cite{takEndtoEndSpectroTemporalGraph2021} and \cite{Jung2021AASIST}. As the amount of phase perturbation increases, the performance decreases on all three CM systems, indicating that time-domain CM systems utilize phase information to discern spoofed speech. As a comparison, Table~\ref{tab:resdegmag} shows the results of all CM systems evaluated with different magnitude perturbation settings. 
\begin{table}[t]
\caption {EER performance \zd{of three} CM systems evaluated \zd{on phase perturbed test data}.}
\label{tab:resdegphase}
{\renewcommand{\arraystretch}{1.2}%
\begin{tabular}{ccccc}
\toprule
\multicolumn{2}{c}{\multirow{2}{*}{EER (\%)}} & \multicolumn{3}{c}{CM Systems} \\
\multicolumn{2}{c}{}                          & RawNet-2  & RawGAT-ST & AASIST \\\toprule
\multicolumn{2}{c}{No Perturbation}            & 4.25      & 1.36      & 1.35   \\\midrule
\multirow{5}{*}{Phase}        & $\frac{1}{2}\pi$         & 9.55      & 20.09     & 21.12  \\
                              & $\pi$           & 19.56     & 29.01     & 48.87  \\
                              & $\frac{3}{2}\pi$         & 27.02     & 31.57     & 56.58  \\
                              & $2\pi$           & 28.02     & 31.82     & 56.91  \\
                              & Pooled        & 21.04     & 28.12     & 45.87  \\\bottomrule
\end{tabular}}
\end{table}


\begin{table}[h]
\caption {EER performance \zd{of three} CM systems evaluated with magnitude \zd{perturbed test data.}}
\label{tab:resdegmag}
{\renewcommand{\arraystretch}{1.2}%
\begin{tabular}{ccccc}
\toprule
\multicolumn{2}{c}{\multirow{2}{*}{EER (\%)}} & \multicolumn{3}{c}{CM Systems}                   \\
\multicolumn{2}{c}{}                          & RawNet-2       & RawGAT-ST      & AASIST         \\\toprule
\multicolumn{2}{c}{No Perturbation}            & 4.25           & 1.36           & 1.35           \\\midrule
\multirow{7}{*}{Mag.}      & 10 dB            & 24.78          & 9.74           & 11.21          \\
                           & 5 dB             & 34.08          & 15.11          & 13.01          \\
                           & 0 dB             & 43.04          & 21.19          & 19.9           \\
                           & -5 dB            & 45.25          & 32.47          & 30.5           \\
                           & -10 dB           & 45.48          & 37.27          & 35.16          \\
                           & Pooled           & 38.53          & 23.16          & 21.96          \\\bottomrule

\end{tabular}}
\end{table}

\begin{figure}[t]
  \centering
  \centerline{\includegraphics[width=0.8\columnwidth]{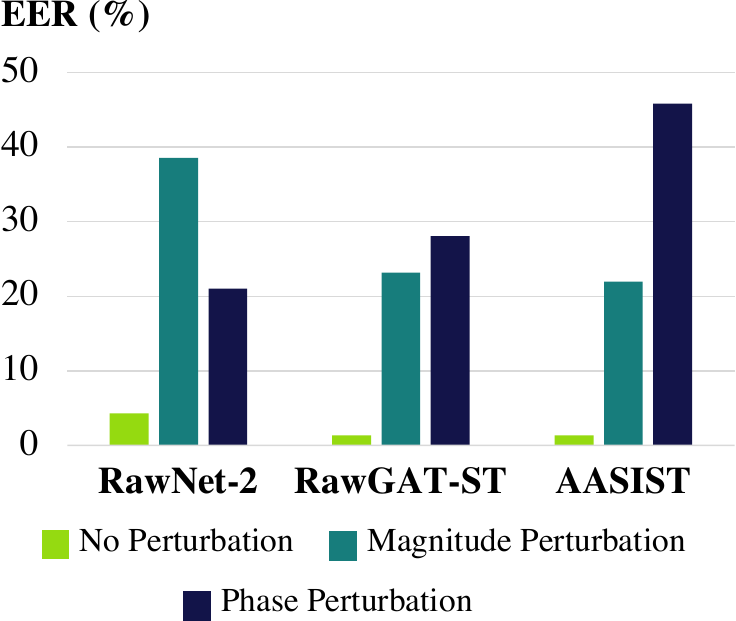}}
\vspace{-1pt}
\caption{EER degradation in phase-perturbed and magnitude-perturbed settings compared to the baseline performance.}
\label{fig:cmtrend}
\end{figure}

We use the pooled EER from all phase or magnitude perturbed settings to represent the performance under this perturbation setting and compare them against the baseline setting. As illustrated in Figure~\ref{fig:cmtrend}, CM systems with better performance in ASVspoof2019LA show more degradation in phase-perturbed settings and less degradation in magnitude-perturbed settings, indicating that better-performing time-domain CM systems also rely more on phase information.

\begin{table*}[]
\centering
\caption{Evaluation results on ASVspoof2021LA. The left side denotes the perturbation setting in training. C1-C7 denote different transmission codecs.}
\label{tab:finalres}
{\renewcommand{\arraystretch}{1.4}%
\begin{tabular}{cc|cccccccc}
\hline
\multicolumn{2}{c|}{{EER (\%)}}                          & C1    & C2    & C3    & C4    & C5    & C6    & C7    & Pooled \\\hline
\multicolumn{2}{c|}{No Perturbation}            & 4.68  & 5.87  & 14.39 & 5.75  & 5.44  & 7.66  & 10.26 & 9.91   \\\hline
\multirow{4}{*}{Phase}           & $\frac{1}{2}\pi$       & 4.49  & 6.18  & 8.68  & 5.18  & 5.80  & 6.35  & 8.20  & 7.33   \\
                                 & $\pi$         & 6.72  & 7.00  & 7.34  & 6.41  & 6.89  & 6.85  & 7.30  & 7.31   \\
                                 & $\frac{3}{2}\pi$       & 5.52  & 6.20  & 10.66 & 5.21  & 6.18  & 7.25  & 6.21  & 8.32   \\
                                 & $2\pi$         & 5.68  & 6.37  & 9.63  & 5.42  & 6.34  & 7.14  & 6.39  & 7.73   \\\hline
\multirow{5}{*}{Magnitude}       & 10 dB       & 7.36  & 8.57  & 19.88 & 8.79  & 8.39  & 9.53  & 14.36 & 14.54  \\
                                 & 5 dB        & 10.40 & 11.46 & 30.87 & 11.96 & 11.35 & 14.45 & 19.09 & 17.80  \\
                                 & 0 dB        & 17.41 & 18.23 & 40.99 & 18.07 & 17.98 & 20.63 & 26.21 & 23.64  \\
                                 & -5 dB       & 23.70 & 25.05 & 46.63 & 24.45 & 25.00 & 29.75 & 34.23 & 29.87  \\
                                 & -10 dB      & 34.77 & 34.55 & 46.84 & 34.49 & 34.95 & 36.55 & 38.84 & 37.40  \\\hline
\end{tabular}}
\end{table*}

\section{Phase perturbation during training improves channel robustness}
In Section 2, our experiments establish the problem of overfitting to phase \zd{information in the trainin data.}.
In this section, we will explore perturbing phase during training to lessen phase reliance for time-domain CM to improve channel robustness.

Communication channels typically employ lossy compression codecs, many of which focus on encoding only frequency-magnitude information, since humans are more sensitive towards them~\cite{painterPerceptualCodingDigital2000}. After transmitting through such compression, \zd{much} phase information is \zd{corrupted, making the performance of time-domain CM systems degrade as they have much reliance on phase information}.
By perturbing the phase during training, we can reduce the CM systems' reliance on phase information to build more robust CM systems. \zd{However, we expect that this perturbation should not be too much that completely removes the reliance on phase, as some useful phase information may still remain for the time-domain CM systems to pick up, even after going through the communication channel.}
We hypothesize that there is a midway setting between not perturbing and perturbing all phases that provides the best trade-off resulting in best performance.

\subsection{Channel-shifted dataset}
To evaluate the performance of CM systems in unseen communication channels, we use ASVspoof2021LA, the logical access (LA) sub-track of the ASVspoof2021 challenge. This subset transmitted the entire ASVspoof2019LA set, along with additional samples, through seven communication channels, denoted as C1 to C7. C1 is the same channel as ASVspoof2019LA, while C2 to C7 are unseen channels. Amongst the unseen channels, C2 and C5 use the time-domain compression algorithms a-law and $\mu$-law, while C4, C6, and C7 employ magnitude-based compression codecs, G.722~\cite{g722}, GSM~\cite{596038}, and OPUS~\cite{valin2012definition}. C3 differs from C2 by transmitting over a public switched telephone network, therefore introducing uncontrollable and unknown artifacts, such as data corruption during transit.

\subsection{Experimental setup}
As \zd{described} in Section 2, all three CM systems heavily utilize phase information. In the interest of saving space while not losing generalization ability, we selected AASIST to study in depth due to its superior performance on ASVspoof2019LA. 

We perturb the phase of the training portion of ASVspoof2019LA with the same four perturbation amounts: $\frac{1}{2}\pi$, $\pi$,$\frac{3}{2}\pi$ and $2\pi$, then evaluate on ASVspoof2021LA. For magnitude, we use the same five magnitude perturbation settings as in Section 2. A baseline setting with no perturbation is also provided. Each setting is trained with three distinct random seeds and results are averaged. To better utilize GPU resources, the batch size of the experiments in this part is slightly increased, while all other hyperparameters 
remain unchanged. 

Following the ASVspoof2021 challenge, when reporting performance on channel-variant data, we use pooled EER of all channels to represent the overall performance of CM systems.

\subsection{Results and analysis}
We begin by confirming our first hypothesis: by perturbing phase or magnitude during training, we can reduce the time-domain CM system's reliance on phase or magnitude. \zd{To do so, we train the CM system with different levels of phase or magnitude perturbation, and test it on phase or magnitude perturbed version of the ASVspoof2019LA evaluation set. For phase perturbed test data, we use the maximum phase perturbation (i.e., $2\pi$), and for magnitude perturbed test data, we use -10 dB SNR for the perturbation.} 
Since \zd{the perturbations are so strong in the test data,} we expect to see performance degradation when there is no perturbation during training, then performance improvement as more phase perturbation is applied. 

\begin{figure}[h]
  \centering
  \includegraphics[width=\columnwidth]{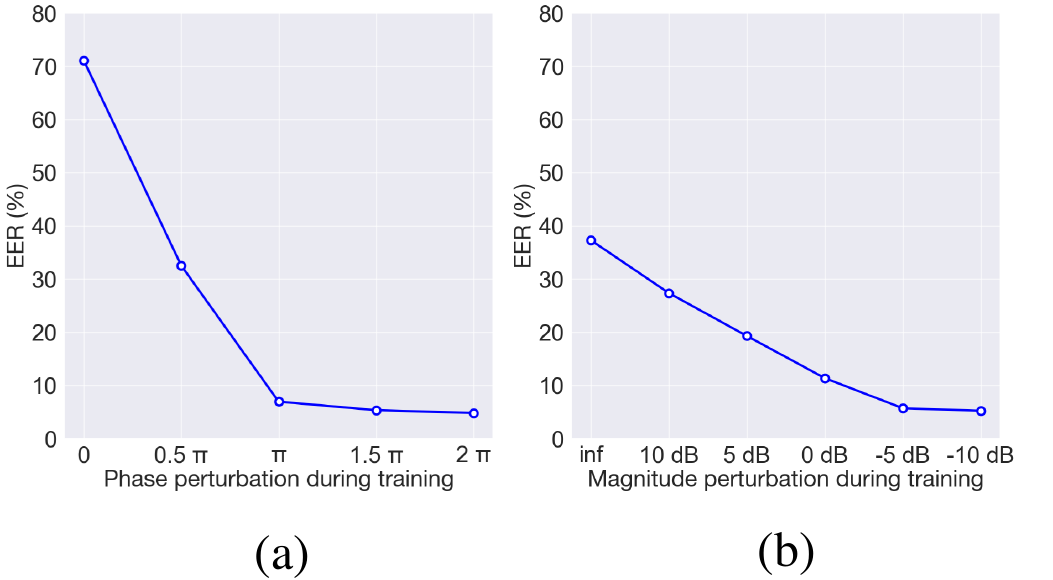}
\vspace{-10pt}
\caption{(a) EER \zd{of AASIST trained with different levels of phase perturbation and evaluated on data with maximum phase perturbation ($2\pi$). (b) EER of AASIST trained with different levels of magnitude perturbation and evaluated on data with -10 dB SNR magnitude perturbation.}}
\label{fig:eer}
\end{figure}

Results are as shown in Figure~\ref{fig:eer}. In Figure~\ref{fig:eer}(a), the performance improves with more phase perturbation in training, indicating less dependence on phase information. Similarly, Figure~\ref{fig:eer}(b) illustrates the results on magnitude \zd{perturbation}. We observe performance improvement as magnitude perturbation increases during training, indicating that the CM system is less dependent on magnitude information.

With this established, we perform evaluation of all training settings on ASVspoof2021LA, and the results are shown in Table~\ref{tab:finalres}. 
As expected, as magnitude is perturbed during training, performance on all unseen communication channels degrades, showing that removing reliance on magnitude information is harmful for CM systems' performance. This suggests that CM systems can benefit from the magnitude information preserved by lossy compression codecs. 

At the same time, all settings with phase perturbed show some improvement in pooled results, \zd{achieving} a more robust overall performance compared to non-perturbed settings. 
This indicates that by being less sensitive to phase information, time-domain CM systems can better generalize to unseen communication channels. Best performing setting shows a \zd{relative EER} improvement of 26.2\% compared to \zd{the no perturbation} baseline, without introducing any channel data during training.
We also notice that the best-performing phase perturbation setting appears at $\pi$, \zd{aligning with} to our hypothesis of a ``midway'' perturbation setting.

Amongst the individual unseen communication channels, C2 and C5 are codecs \zd{encoding the time-domain waveform directly}, and we observe slight performance degradation as phase information is perturbed. C4, C6, and C7, \zd{on the other hand, uses encode magnitude spectrogram with lossy compression, and} 
we see performance improvement with phase perturbation. Interestingly, we notice that on different channels, the best-performing phase perturbation setting appears at different perturbation amounts. This \zd{suggests that different amount of useful phase information is retained by different compression codecs.} 

C3 is transmitted through real-world phone lines and has more unknown data corruption en route, and we observe that phase perturbation during training \zd{improves} performance. At different phase perturbation settings, C3 also shows more fluctuation, indicating that transmission artifacts are more likely to introduce phase artifacts as well.

On the C1 condition, even though no unseen communication artifacts are present, we still see a slight performance improvement at $\frac{1}{2}\pi$. We believe \zd{that} this is likely \zd{because} phase perturbation also brings better generalization ability to unseen attacks by masking part of model-specific artifacts. Speech generative systems typically take magnitude features as input and synthesize time-domain data; therefore, they have a heavier burden generating phase information. This makes speech generative algorithms more prone to creating phase artifacts that are model-specific. As we have established time-domain CM systems' reliance on phase information, we believe CM systems may have overfitted to specific artifacts in training data, and phase perturbation can also aid in mitigating this effect.

\section{Conclusions}
In this paper, we observed significant degradation of various state-of-the-art time-domain CM systems when evaluated on phase-perturbed speech utterances. \zd{This degradation may cause channel robustness issues}, since many communication channels employ lossy codecs that only encode frequency-magnitude information \zd{while losing much phase information}. We proposed to mitigate this \zd{issue} by perturbing phase in training. \zd{Systematic evaluation on real-world} channel-variant data verified that perturbing phase in training \zd{does significantly improve the channel robustness of a state-of-the-art} time-domain CM system.
%
\zd{For future work, we plan to use this insight to design CM systems that strike the balance between modeling useful phase information and being less sensitive to channel phase alternation.}

\section{Acknowledgement}
This work is partially supported by a New York State Center of Excellence in Data Science award, and synergistic activities funded by the National Science Foundation (NSF) under grant DGE-1922591.

\vfill\pagebreak

\bibliographystyle{IEEEtran}
\bibliography{mybib}

\end{document}